\documentclass[11pt, reqno]{amsart}
\usepackage{graphics,color}
\usepackage{amsfonts}

\def\Box{\vcenter{\vbox{\hrule\hbox{\vrule
     \vbox to 8.8pt{\hbox to 10pt{}\vfill}\vrule}\hrule}}}

\newtheorem{theorem}{Theorem}
\newtheorem{lemma}{Lemma}

\newtheorem{proposition}{Proposition}

\newtheorem{remark}{Remark}

\newcommand{\tr}{\mathrm{tr}}

\newcommand{\F}{\mathbf{F}}

\def\tr{\operatorname{tr}}

\numberwithin{equation}{section}

\begin{document}

\title[]
 {Period-Different $m$-Sequences With At Most A Four-Valued Cross Correlation}

\author{Lei Hu}
\author{Xiangyong Zeng}
\author{Nian Li}
\author{Wenfeng Jiang}

\address{Lei Hu and Wenfneg Jiang are with the State Key Laboratory of Information Security, Graduate
University of Chinese Academy of Sciences, Beijing, 100049, China.
Email: hu@is.ac.cn}
\address{Xiangyong Zeng and Nian Li are with the Faculty of Mathematics and Computer Science,
Hubei University, Wuhan,430062, China. Email: xzeng@hubu.edu.cn}

\keywords{$m$-sequence, cross correlation, quadratic form}

\date{}

\begin{abstract}
In this paper, we follow the recent work of Helleseth, Kholosha,
Johanssen and Ness to study the cross correlation between an
$m$-sequence of period $2^m-1$ and the $d$-decimation of an
$m$-sequence of shorter period $2^{n}-1$ for an even number $m=2n$.
Assuming that $d$ satisfies $d(2^l+1)=2^i({\rm mod}\,\, 2^n-1)$ for
some $l$ and $i$, we prove the cross correlation takes exactly
either three or four values, depending on ${\rm gcd}(l,n)$ is equal
to or larger than 1. The distribution of the correlation values is
also completely determined. Our result confirms the numerical
phenomenon Helleseth et al found. It is conjectured that there are
no more other cases of $d$ that give at most a four-valued cross
correlation apart from the ones proved here.
\end{abstract}
\maketitle

\section{Introduction}

Sequences with good correlation properties have important
applications in communication systems. The maximal period sequences
($m$-sequences) and their decimations are widely used to design
sequence families with low-correlation \cite{GG,H76,HK,Rosendahl}.

Recently, Ness and Helleseth initiated the studies on the cross
correlation between an $m$-sequence $\{s_t\}$ of period $2^m-1$ and
the $d$-decimation $\{u_{dt}\}$ of an $m$-sequence $\{u_{t}\}$ with
shorter period $2^{n}-1$ for $m=2n$ and $d$ with ${\rm
gcd}(d,2^{n}-1)=1$ \cite{NH4}. These two period-different sequences
are exactly the $m$-sequences used to construct the Kasami sequence
family \cite{Kasami,Kasami69,ZLH}. They proposed the first family of
decimation $d$ that the cross correlation takes three values in
\cite{NH4}, where $n$ is odd and $d$ is taken as $d={2^{n}+1\over
3}$. They also completed a full search for all the decimation $d$
giving at most a six-valued correlation for $m\leq 26$ \cite{NH4}.
Later, they found another family of $d=2^{n+1\over 2}-1$ giving a
three-valued cross correlation for odd $n$ \cite{NH10}. Further, in
\cite{HKN6}, Helleseth, Kholosha, and Ness showed for a larger
family of $d$ satisfying
\begin{equation}\label{d0}d(2^l+1)=2^i\,({\rm mod}\, 2^{n}-1),\end{equation}
the cross correlation is three-valued, where $n$ is odd and ${\rm
gcd}(l,n)=1$. Based on their numerical experiment, the authors in
\cite{HKN6} conjectured that they had found all $d$ giving a
three-valued cross correlation.

For the cross correlation taking four values, Ness and Helleseth
\cite{NH11} proved this is the case if $d=\frac{2^{3k}+1}{2^k+1}$
and $m=6k$. This result was very recently generalized to the case of
$d=\frac{2^{kr}+1}{2^k+1}$ and $m=2kr$ with odd $r$ by Helleseth,
Kholosha and Johanssen in \cite{HKJ}. As they had pointed out, there
still existed  some examples of four-valued cross correlation that
did not fit into all the known families they had found.

In this paper, we follow above work 
to study the cross correlation of $\{s_t\}$ and $\{u_{dt}\}$ with
the decimation $d$ satisfying Equality (\ref{d0}) for a general case
of parameters $m$ and $l$, namely, without any restriction on $n$
and $l$. We prove the cross correlation takes exactly either three
or four values, depending on ${\gcd}(l,n)$ is equal to or larger
than 1. The distribution of the correlation values is also
completely determined.

Our result theoretically confirms the numerical experiments
Helleseth et al done in a unified way, that is, except some
exceptional $d$ in small $m=8$ and 10, all $d$ found in the
experiments and giving at most a four-valued cross correlation are
covered by our result. 
This definitely includes new classes of $d$ that can not be
explained in previous results, for instance, $d=181$ for $m=20$. The
full numerical search of up to $m=26$ \cite{NH4} together with our
result suggests one to conjecture that for $m>10$ there are no more
other $d$ giving at most a four-valued cross correlation apart from
the ones satisfying (1.1).


Our proof is mainly based on the theory of quadratic forms and on
the study of a class of equations over a finite field, which is a
little different from the treatments of Helleseth et al.

This paper is organized as follows. Section 2 is preliminaries on
cross correlation and quadratic forms. Section 3 determines the
ranks of quadratic polynomials $\rho_a(x)$ and the enumeration of
the ranks as $a$ varies in $\F_{2^n}$. In Section 4, we obtain the
desired four-valued cross correlation distribution. 

\section{Preliminaries on Cross Correlation and Quadratic Forms}

Let $\F_{2^m}$ be the finite field with $2^m$ elements and
$\F_{2^m}^*=\F_{2^m}\setminus\{0\}$. For any integers $m$ and $n$
with $m\,|\,n$, the trace function from $\F_{2^m}$ to $\F_{2^n}$ is
defined by $\tr^m_n(x)=\sum\limits_{i=0}^{m/n-1}x^{2^{ni}}$, where
$x$ is an element in $\F_{2^m}$.

Let $m=2n$ be an even integer, $\alpha$ be a primitive element of
$\F_{2^m}$ and $T=2^n+1$. Then $\beta=\alpha^T$ is a primitive
element of $\F_{2^n}$. The $m$-sequence $\{s_t\}$ of period $2^m-1$
and the $m$-sequence $\{u_t\}$ of shorter period $2^n-1$ are given
by
$$s_t=\tr_1^m(\alpha^t),\,\,\,\,u_t=\tr_1^n(\beta^t),$$ respectively.
The cross correlation at shift $0\leq \tau\leq 2^n-2$ between
$\{s_t\}$ and the $d$ decimation of $\{u_{t}\}$  is defined by
$$C_d(\tau)=\sum_{t=0}^{2^m-2}(-1)^{s_t+u_{d(t+\tau)}}.$$

In this paper, we consider the cross correlation for the decimation
$d$ satisfying
\begin{equation}\label{d}d(2^l+1)=2^i\,({\rm mod\,} 2^n-1)\end{equation}
for some $l$ and $i$ for an arbitrary $n$. Note that in \cite{HKN6},
only the case for odd $n$ and $\gcd(l,n)=1$ is considered.
Obviously, if $d$ satisfies Equality (\ref{d}), then both $d$ and
$2^l+1$ are prime to $2^n-1$. In fact, for each $d$ satisfying
(\ref{d}), one can choose some $l<n$ such that both (\ref{d}) and
$\gcd(2^l+1,2^m-1)=1$ hold. To explain this, we need the following
standard consequence of the division algorithm.

\begin{lemma}\label{lem2}  Let $u$ and $u$
be positive integers with $w={\rm gcd}(u,v)$, then $${\rm
gcd}(2^u-1,2^v-1)=2^w-1$$ and $${\rm gcd}(2^u-1,2^v+1)=
\left\{\begin{array}{ll}1,&{\rm if}\,\, {u}/{w}\,\, {\rm is\,\,
odd,}\\1+2^w,&{\rm otherwise.}
\end{array}\right.$$
\end{lemma}

Assume that Equality (\ref{d}) holds. Write $n=2^{e}n_1$ with odd
$n_1$. By Lemma \ref{lem2}, $2^e$ divides $l$, and $l$ can be
written as $l=2^el_1$. If $l_1$ is even, then by Lemma \ref{lem2}
again, $\gcd(2^l+1,2^m-1)=1$. Otherwise, if $l_1$ is odd, let
$l'=n-l$. Then one has $d(2^{l'}+1)=2^{n-l+i}\,({\rm mod\,} 2^n-1)$
and $\gcd(2^{l'}+1,2^m-1)=1$.

This leads we use the following notation on integers:
\begin{itemize}

\item $n=kr$ and $m=2kr$ with an arbitrary integer $k$ and odd $r\geq 3$.


\item $l=ks$, where $0<s<r$, $s$ is even, and gcd$(r,s)=1$. Note
that $\gcd(2^l+1,2^m-1)=1$ by Lemma \ref{lem2}.

\item $d$ is a decimation satisfying $d(2^l+1)=2^i\,({\rm mod}\, 2^n-1)$ for some $i$.

\end{itemize}

Additionally, we let

\begin{itemize}

\item $N(f,\F_{2^u})$ denote the number of roots in a finite field $\F_{2^u}$ of a polynomial $f$.

\end{itemize}

For any function $f(x)$ from $\F_{2^m}$ to $\F_2$, the trace
transform $\widehat{f}(\lambda)$ of $f(x)$ is defined by
$$\widehat{f}(\lambda)=\sum\limits_{ x\in
\F_{2^m}}(-1)^{f(x)+\tr_1^m(\lambda x)},\,\,\,\lambda\in
\F_{2^m}.$$ 
$f(x)$ is called to be a quadratic form, if it can be expressed as a
degree two polynomial of the form
$f(x_1,x_2,\cdots,x_m)=\sum\limits_{1\leq i\leq j\leq
m}a_{ij}x_ix_j$ by using a basis of $\F_{2^m}$ over $\F_2$.
The rank of the symmetric matrix with zero diagonal entries and
$a_{ij}$ as the $(i,j)$ and $(j,i)$ entries for $i\neq j$, is
defined as the rank of $f(x)$. It can be calculated from the number
$N$ of $x\in \F_{2^m}$ such that $B_f(x,z):=f(x+z)+f(x)+f(z)=0$
holds for all $z\in \F_{2^m}$, namely, $\label{rank0}{\rm
rank}(f)=m-{\rm log}_2(N)$. The $B_f(x,z)$ is called the symplectic
form of $f(x)$. It is known that the distribution of trace transform
values of a quadratic form is determined by its rank as the
following lemma. For more details, the reader is referred to
\cite{HK}.

\begin{lemma}[\cite{HK}]\label{lem1} The rank of any quadratic
form is even. Let $f(x)$ be a quadratic form on $\F_{2^m}$ with rank
$2h$, $1\leq h\leq m/2$. Then its trace transform values have the
following distribution:
$$\widehat{f}(\lambda)= \left\{\begin{array}{ll}\pm2^{m-h}
,&2^{2h-1}\pm 2^{h-1} \,\,\,{\rm times,}\\0,&2^m-2^{2h} \,\,\,{\rm
times.}
\end{array}\right.$$
\end{lemma}

Using the trace representation of the sequences, $C_d(\tau)$ can be
written as
$$\begin{array}{rcl}C_d(\tau)
&=&\sum\limits_{t=0}^{2^m-2}(-1)^{s_t+u_{d(t+\tau)}}\\
&=&\sum\limits_{t=0}^{2^m-2}(-1)^{\tr_1^m(\alpha^t)+\tr_1^n(\beta^{d\tau}\alpha^{tdT})}\\
&=&\sum\limits_{x\in \F_{2^m}^*}(-1)^{\tr_1^m(x)+\tr_1^n(ax^{dT})},
\end{array}$$
where $a=\beta^{d\tau}\in \F_{2^n}$. Now making a substitution
$x=y^{2^l+1}$, we have
\begin{equation}\label{rho1}\begin{array}{rcl}C_d(\tau)
&=&\sum\limits_{y\in
\F_{2^m}^*}(-1)^{\tr_1^m(y^{2^{l}+1})+\tr_1^n(ay^{2^n+1})}.
\\&=&\sum\limits_{y\in
\F_{2^m}^*}(-1)^{\rho_a(x)}=-1+\widehat{\rho_a}(0),
\end{array}\end{equation}
where
\begin{equation}\label{rho}\rho_a(x)=\tr_1^m(x^{2^{l}+1})+\tr_1^n(ax^{2^n+1}).\end{equation}

The idea of the above work handling $C_d(\tau)$ comes from Ness and
Helleseth \cite{NH4}. However, the observation that
$\gcd(2^l+1,2^m-1)=1$ and the substitution $x=y^{2^l+1}$ is
one-to-one over $\F_{2^m}$ enables us to simplify the calculation of
$C_d(\tau)$, comparing with the work of \cite{NH4}.


\section{Ranks of Quadratic Forms $\rho_a(x)$ and Their Enumeration}

In this section, we determine the ranks of the quadratic forms
$\rho_a(x)$ defined by Equality (\ref{rho}), and the enumeration of
the ranks when $a$ ranges over $\F_{2^n}^*$.

The symplectic form of $\rho_a(x)$ is given by
$$\begin{array}{rcl}B_{\rho_a}(x,z)&=&\rho_a(x)+\rho_a(z)+\rho_a(x+z)\\
&=&\tr_1^m(zx^{2^{l}}+xz^{2^{l}})+\tr_1^n(azx^{2^n}+axz^{2^n})
\\&=&\tr_1^m(zx^{2^{l}}+zx^{2^{m-l}})+\tr_1^n(\tr_n^m(azx^{2^n}))
\\&=&\tr _1^m(z(x^{2^{l}}+x^{2^{m-l}}+ax^{2^n}))\\
&=&\tr_1^m(z^{2^{l}}(x^{2^{2t}}+a^{2^l}x^{2^{t}}+x)),
\end{array}$$
where $t=n+l$.

Therefore, we consider the number of roots in $\F_{2^m}$ of the
polynomial
\begin{equation}\label{equation1}f_a(x):=x^{2^{2t}}+a^{2^l}x^{2^{t}}+x
,\,\,a\in \F_{2^n}^*.\end{equation} Note that $\gcd(t,m)=k$, all
these roots form an $\F_{2^k}$-vector space. Thus, the number of
roots of $f_a(x)$ is a power of $2^k$.

Denote $y=x^{2^t-1}$. We have
\begin{equation}\label{equation6}f_a(x)=x(y^{2^{t}+1}+a^{2^l}y+1).
\end{equation}
The nonzero roots of $f_a(x)$ are closely related to that of the
polynomial
\begin{equation}\label{equation7}g_a(y):=y^{2^t+1}+a^{2^l}y+1,\end{equation}
or equivalently, to that of the roots of the polynomial
\begin{equation}\label{equation7'}h_c(z):=z^{2^t+1}+cz+c,\end{equation}
which is obtained from Equation (\ref{equation7}) by substituting
$y=a^{-2^l}z$, then dividing by $a^{-2^{l}(2^t+1)}$ and letting
$c=a^{2^{l}(2^t+1)}$. Notice that there exists a one-to-one
correspondence between $a\in \F_{2^n}^*$ and $c\in \F_{2^n}^*$ since
$\gcd(2^t+1,2^n-1)=1$.

The properties about the roots of a polynomial with the form as
Equation (\ref{equation7'}) were investigated technically in \cite
{B} and \cite{HK07}. To facilitate the introduction of their
results, we still use the notation $h_c(x)$ defined by Equality
(\ref{equation7'}) in the following Lemmas \ref{lem3}. But notice
that this lemma holds for any positive integers $m$ and $t$, and
$c\in\F_{2^m}$.

\begin{lemma}[Theorem 5.4 of \cite{B}]\label{lem3} Denote
\begin{equation}\label{xi}
\Xi_m=\{\xi\in
\widetilde{\F}_{2^m}\,|\begin{array}{l}\,\xi^{2^t-1}=\frac{1}{\eta+1},\,\,{\rm
where}\,\,\eta\in\widetilde{\F}_{2^m}\,\, {\rm
and}\,\,h_c(\eta)=0\end{array}\},\end{equation} where
$\widetilde{\F}_{2^m}$ denotes the algebraic closure of $\F_{2^m}$.
 Let $k={\rm gcd}(t,m)$.
Then $h_c(x)$ has either $0$, $1$, $2$ or $2^k+1$ roots in
$\F_{2^m}$. Moreover, if $h(x)$ has exactly one root in $\F_{2^m}$,
then $\Xi_m\cap \F_{2^m}$ is nonempty, and $\tr ^m_{k}(\xi)\neq 0$
for any $\xi\in \Xi_m\cap \F_{2^m}$.
\end{lemma}

\begin{proposition}\label{pro1} For any $a\in \F_{2^n}^*$, $g_a(y)$ has
either $0$, $2$, or $2^k+1$ nonzero roots in $\F_{2^m}$.
\end{proposition}

 \begin{proof} Since $g_a(y)$ has the same number of roots as $h_c(z)$ for
$c=a^{2^{t}(2^t+1)}$, it is sufficient to prove $h_c(z)$ has either
$0$, $2$, or $2^k+1$ nonzero roots in $\F_{2^m}$.

By $\gcd(t,n)=k$ and Lemma \ref{lem3}, $h_c(z)$ has either $0$, $1$,
$2$, or $2^k+1$ nonzero roots in $\F_{2^m}$. Thus, we only need to
prove $h_c(z)$ can not have exactly one root in $\F_{2^m}$.

Assume that $h_c(z)$ has exactly one root in $\F_{2^m}$, denoted by
$\eta$. Then $\eta^{2^n}$ also is a root of $h_c(z)$, which implies
that $\eta=\eta^{2^n}$ and then $\eta\in \F_{2^n}$. Thus, $h_c(z)$
has exactly one root in $\F_{2^n}$ too. By Lemma \ref{lem3}, the set
$\Xi_n\cap \F_{2^n}\subseteq \Xi_m\cap \F_{2^m}$ is nonempty, and
for any $\xi\in \Xi_n\cap \F_{2^n}\subseteq \Xi_m\cap \F_{2^m}$, we
have $\tr^m_k(\xi)\neq 0$, which contradicts with the fact
$\tr^m_k(\xi)=\tr^n_k(\tr^m_n(\xi)))=\tr^n_k(\xi\cdot
\tr^m_n(1))=0$.


\end{proof}

\begin{proposition}\label{pro2}\label{pro2} Let $a\in \F_{2^n}^*$. Then

 (1) If $y_1$ and $y_2$ are two different roots of $g_a(y)$ in $
\F_{2^m}$, then $y_1y_2$ is a $(2^k-1)$-th power in $ \F_{2^m}$; and

 (2) If there exist at least three different roots of $g_a(y)$ in $
\F_{2^m}$, then each root of $g_a(y)$ in $ \F_{2^m}$ is $(2^k-1)$-th
power in $ \F_{2^m}$.
\end{proposition}

\begin{proof} (1) If $g_c(y_i)=0$ for $i=1$ and $2$, then we have
$$\begin{array}{rcl}y_1y_2(y_1+y_2)^{2^{t}}&=&y_1^{2^{t}+1}y_2+y_1y_2^{2^{t}+1}\\
&=&(a^{2^l}y_1+1)y_2+(a^{2^l}y_2+1)y_1\\
&=&y_1+y_2.\end{array}$$ Since $\gcd(2^{t}-1,2^m-1)=2^k-1$, so
$$y_1y_2=(y_1+y_2)^{-(2^{t}-1)}$$
is  a $(2^k-1)$-th power in $ \F_{2^m}$.

(2) Let $y_1, y_2$ and $y_3$ be three different roots of $g_a(y)$ in
$ \F_{2^m}$. Then by (1), all of $y_1y_2,y_1y_3$ and $y_2y_3$ are
$(2^k-1)$-th power in $ \F_{2^m}$, and hence so are
$y_1^2=(y_1y_2)(y_1y_3)/(y_2y_3)$ and $y_1$. This finishes the
proof.
\end{proof}

\begin{proposition}\label{pro3} Let $a\in \F_{2^n}^*$. Then $f_a(x)$ has exactly
either 1 or $2^{2k}$ roots in $\F_{2^m}$, that is, $\rho_a(x)$ has
rank of $m$ or $m-2k$. Furthermore,
$$N(f_a,\F_{2^m})=1 \Longleftrightarrow  N(g_a,\F_{2^m})=0 {\rm\,\, or \,\,}2 $$
and
$$N(f_a,\F_{2^m})=2^{2k} \Longleftrightarrow  N(g_a,\F_{2^m})=2^k+1. $$
\end{proposition}

\begin{proof} This is shown by analyzing the number of roots in
$\F_{2^m}$ of $g_a(y)$. 
Note that $y=x^{2^t-1}$ and $\gcd(2^m-1,2^t-1)=2^k-1$. Thus, the
correspondence from $x\in \F_{2^m}^*$ to $y$ is $(2^k-1)$-to $1$.

If $g_a(y)$ has exactly $2^k+1$ roots $y$ in $\F_{2^m}$, then by
Proposition \ref{pro2}(2), all its roots in $\F_{2^m}$ are
$(2^k-1)$-th power in $ \F_{2^m}$, and they correspond
$(2^k+1)(2^k-1)=2^{2k}-1$ nonzero roots of $f_a(x)$ under the
inverse of the mapping $x\mapsto y=x^{2^{t}-1}$. Thus $f_a(x)$  has
exactly $2^{2k}$ roots in $\F_{2^m}$.

If $g_a(y)$ has exactly 2 roots $y_1,y_2\in \F_{2^m}$, then by
Proposition \ref{pro2}(1), $y_1,y_2$ are either both $(2^k-1)$-th
powers in $ \F_{2^m}$ or both not $(2^k-1)$-th power in $ \F_{2^m}$.
The former can not be true. If it is that case, then $y_1$ and $y_2$
would correspond total $2(2^k-1)=2^{k+1}-2$ nonzero roots of
$f_a(x)$, and thus $f_a(x)$ would have exactly $2^{k+1}-1$ roots in
$\F_{2^m}$. This is a contradiction with that the number of roots is
a power of $2^k$. Therefore, both two roots are not $(2^k-1)$-th
powers in $ \F_{2^m}$ and $f_a(x)$ has no nonzero root in
$\F_{2^m}$.
\end{proof}

In the sequel, we determine how many $a\in \F_{2^n}$ there are such
that $f_a(x)$ has exactly 1 or $2^{2k}$ roots in $\F_{2^m}$. Note
that we encounter here two different fields $\F_{2^n}$ and
$\F_{2^m}$. To this end, we first have

\begin{proposition}\label{pro4}
$N(g_a,\F_{2^m})= N(g_a,\F_{2^n}) ({\rm\,mod\,}2)$.
\end{proposition}

\begin{proof} Since $g_a(y)$ is a polynomial over $\F_{2^n}$, we
know that for any root $\gamma\in \F_{2^m}$ of $g_a(y)$, its
$2^n$-th power $\gamma^{2^n}$ is also a root of $g_a(y)$ in
$\F_{2^m}$. If this root $\gamma\notin \F_{2^n}$, then $\gamma\neq
\gamma^{2^n}$, and these two elements form a pair
$\{\gamma,\gamma^{2^n}\}$ as
roots of $g_a(y)$ in $\F_{2^m}$, 
By $(\gamma^{2^n})^{2^n}=\gamma$, every two pairs of roots of this
form are either just the same one or disjointed. Thus,
$N(g_a,\F_{2^m})= N(g_a,\F_{2^n}) ({\rm\,mod\,} 2)$.
\end{proof}

To determine $N(f_a,\F_{2^n})$, we need the following lemma. Notice
that in this lemma, integers $n$ and $t$ refer to any positive
integers, and $c$ refers to any element in $\F_{2^n}$.

\begin{lemma}[Theorem 5.6 of \cite{B}]\label{lem5}  Let
$k={\rm gcd}(t,n)$ and $q=2^{k}$, then $h_c(x)$ has either $0$, $1$,
$2$ or $q+1$ roots in $\F_{2^n}$. Moreover, let $N_i$ denote the
number of $c\in \F^*_{2^n}$ such that $h_c(x)$ has exactly $i$ roots
in $\F_{2^n}$, where $i=0$, $1$, $2$, $q+1$. If $\mu=n/k$ is odd,
then
$$\begin{array}{l}
N_0=\frac{q^{\mu+1}+q}{2(q+1)},\,\,\,N_1=q^{\mu-1}-1,\,\,\,
N_2=\frac{(q-2)(q^\mu-1)}{2(q-1)},\,\,\,N_{q+1}=\frac{q^{\mu-1}-1}{q^2-1}.
\end{array}$$
\end{lemma}

\vspace {2mm}

\begin{proposition}\label{pro5} For any $a\in \F_{2^n}^*$, the rank of
$\rho_a(x)$ is $m$ or $m-2k$. Furthermore, let $R_m$ ($R_{m-2k}$,
respectively) be the number of $a\in \F_{2^n}^* $ such that $ {\rm
rank}(f_a)=m$ (${\rm rank}(f_a)=m-2k$, respectively), then
$$R_m=\frac{2^{n+2k}-2^{n+k}-2^n+1}{2^{2k}-1},\,\,
R_{m-2k}=\frac{2^{n+k}-2^{2k}}{2^{2k}-1}.$$
\end{proposition}

\begin{proof} By Propositions \ref{pro1}, \ref{pro3} and \ref{pro4},
we have
\begin{equation}\label{rank1}\begin{array}{l}N(f_a,\F_{2^m})=1
\Longleftrightarrow N(g_a,\F_{2^m})=0 {\rm\,\, or
\,\,}2\Longleftrightarrow N(g_a,\F_{2^n})=0 {\rm\,\, or \,\,}2
\end{array}
\end{equation} and
\begin{equation}\label{rank2}\begin{array}{l}N(f_a,\F_{2^m})=2^{2k}
\Longleftrightarrow  N(g_a,\F_{2^m})=2^k+1 \Longleftrightarrow
N(g_a,\F_{2^n})=1 {\rm\,\, or\,\,}2^k+1.
\end{array}
\end{equation}

Since $\gcd(t,n)=k$ and ${n\over\gcd(t,n)}=r$ is odd, by Lemma
\ref{lem5}, the number of roots of $h_c(z)$ is equal to either $0$,
$1$, $2$ or $2^k+1$. For $i=0$, $1$, $2$ and $2^k+1$, let $N_i$
denotes the number of $c\in \F^*_{2^n}$ such that $h_c(z)$ has
exactly $i$ roots in $\F_{2^n}$. Then
$$\begin{array}{l}
N_0=\frac{2^{k(r+1)}+2^k}{2(2^k+1)},\,\,\, N_1=2^{k(r-1)}-1,\,\,\,
N_2=\frac{(2^k-2)(2^{kr}-1)}{2(2^k-1)},\,\,\,N_{2^k+1}=\frac{2^{k(r-1)}-1}{2^{2k}-1}.
\end{array}$$
Since $g_a(y)$ has the same number of roots as $h_c(z)$ for
$c=a^{2^{l}(2^t+1)}\in \F_{2^n}^*$, and the correspondence from
$c\in\F_{2^n}^*$ to $a\in\F_{2^n}^*$ is bijective by
$\gcd(2^t+1,2^n-1))=1$, so there are $N_i$ elements $a\in
\F_{2^n}^*$ such $g_a(y)$ has exactly $i$ roots in $\F_{2^n}$, which
implies that there are $N_0+N_2$ elements $a\in\F_{2^n}^*$ such that
$f_a(x)$ has exactly one root in $\F_{2^m}$ by Equality
(\ref{rank1}). Then we have
$$\begin{array}{l}R_m=N_0+N_2
=\frac{2^{k(r+1)}+2^k}{2(2^k+1)}+\frac{(2^k-2)(2^{kr}-1)}{2(2^k-1)}
=\frac{2^{n+2k}-2^{n+k}-2^n+1}{2^{2k}-1},\end{array}$$ and
$$\begin{array}{c}R_{m-2k}=N_1+N_{2^k+1}=\frac{2^{n+k}-2^{2k}}{2^{2k}-1}.
\end{array}$$

\end{proof}

\section{Distribution of the Cross Correlation $C_d(\tau)$}
\vspace{2mm}
\begin{lemma}[\cite{NH4}]\label{lem6} Let $m=2n$, then for any decimation $d$
with ${\rm gcd}(d,2^n-1)=1$, the cross correlation value $C_d(\tau)$
satisfies the following relations:

(1) $\sum\limits_{\tau=0}^{2^n-2}C_d(\tau)=1$;

(2) $\sum\limits_{\tau=0}^{2^n-2}(C_d(\tau)+1)^2=2^m(2^n-1)$;

(3)$\sum\limits_{\tau=0}^{2^n-2}(C_d(\tau)+1)^3=-2^{2m}+(\nu+3)2^{n+m}$,
where $\nu$ is the number of solutions $(x_1,x_2)\in
(\F_{2^m}^*,\F_{2^m}^*)$ of the equations
$$\left\{\begin{array}{lc}
x_1+x_2+1=0, \\
x_1^{d(2^n+1)}+x_2^{d(2^n+1)}+1=0.
\end{array}\right.$$
\end{lemma}

\begin{proposition}\label{pro6} Let $d$ be defined by Equality (\ref{d}).
Then there are $2^k-2$ solutions $(x_1,x_2)\in
(\F_{2^m}^*,\F_{2^m}^*)$
 of the system of equations
\begin{equation}\label{equation8}\left\{\begin{array}{lc}
x_1+x_2+1=0, \\
x_1^{d(2^n+1)}+x_2^{d(2^n+1)}+1=0.
\end{array}\right. \end{equation}
\end{proposition}

\begin{proof} Suppose that $x_1,x_2\in \F_{2^m}^*$ and let
$x_1=y_1^{2^{l}+1}$, $x_2=y_1^{2^{l}+1}$. Since ${\rm
gcd}(2^m-1,2^{l}+1)=1$, so there exists a one-to-one correspondence
between $(x_1,x_2)\in (\F_{2^m}^*,\F_{2^m}^*)$ and $(y_1,y_2)\in
(\F_{2^m}^*,\F_{2^m}^*)$. Then Equation (\ref{equation8})
equivalently becomes
$$\left\{\begin{array}{lc}
y_1^{2^{l}+1}+y_2^{2^{l}+1}+1=0, \\
y_1^{d(2^{l}+1)(2^n+1)}+y_2^{d(2^{l}+1)(2^n+1)}+1=0.
\end{array}\right. $$
Since $d(2^{l}+1)\equiv 2^i ({\rm mod}\,2^n-1)$, we have
$$\begin{array}{l}
 y_1^{d(2^{l}+1)(2^n+1)}+y_2^{d(2^{l}+1)(2^n+1)}+1
=y_1^{2^i(2^n+1)}+y_2^{2^i(2^n+1)}+1
=(y_1^{2^n+1}+y_2^{2^n+1}+1)^{2^i}=0.\end{array}$$ Thus, the above
system of equations becomes
\begin{equation}\label{equation9}\left\{\begin{array}{lc}
y_1^{2^{l}+1}+y_2^{2^{l}+1}+1=0, \\
y_1^{2^n+1}+y_2^{2^n+1}+1=0,
\end{array}\right. \end{equation}
which implies
$$\begin{array}{l}y_1^{(2^n+1)(2^{l}+1)}
=(y_2^{2^{m}+1}+1)^{2^n+1}
=y_2^{(2^n+1)(2^{l}+1)}+y_2^{(2^{l}+1)2^n}+y_2^{2^{l}+1}+1
\end{array}$$
and
$$\begin{array}{l}y_1^{(2^n+1)(2^{l}+1)}
=(y_2^{2^n+1}+1)^{2^{l}+1}=y_2^{(2^n+1)(2^{l}+1)}
+y_2^{(2^n+1)2^{l}}+y_2^{2^n+1}+1.
\end{array}$$

Then we have
$$\begin{array}{lcl}y_2^{(2^{l}+1)2^n}+y_2^{2^{l}+1}+y_2^{(2^n+1)2^{l}}
+y_2^{2^n+1}
&=&(y_2^{2^{n+l}}+y_2)(y_2^{2^n}+y_2^{2^{l}})\\
&=&(y_2^{2^{n+l}}+y_2)(y_2^{2^{n-l}}+y_2)^{2^{l}}\\
&=&0.\end{array}$$ So, $y_2$ belongs to $\F_{2^{n+l}}\bigcap
\F_{2^{m}}=\F_{2^k}$ or $\F_{2^{n-l}}\bigcap \F_{2^{m}}=\F_{2^k}$.
In either case, we always have $y_2\in \F_{2^k}$. Similarly, we have
$y_1\in \F_{2^k}$. Then, Equation (\ref{equation9}) becomes
\begin{equation}\label{equation10}
y_1^{2}+y_2^{2}+1=0,
\end{equation}
or equivalently,
\begin{equation}\label{equation10}
y_1+y_2+1=0
\end{equation} with $y_1,y_2\in \F_{2^k}^*$, which has exactly
$2^k-2$ solutions. Therefore, Equation (\ref{equation8}) has $2^k-2$
solutions $(x_1,x_2)\in (\F_{2^m}^*,\F_{2^m}^*)$. This finishes the
proof.
\end{proof}

\begin{theorem}\label{thm1} The cross-correlation function $C_d(\tau)$ has the
following distribution:

$$\left\{\begin{array}{ll}
-1,&{\rm occurs}\,2^{n-k}-1\,\,\,{\rm times,}\\
-1+2^{n},&{\rm occurs}\,\frac{(2^{n}+1)2^{k-1}}{2^{k}+1}\,\,\,{\rm
times,}\\
-1-2^{n},&{\rm occurs}\,\frac{(2^n-1)(2^{k-1}-1)}{2^{k}-1}\,\,\,{\rm
times,}\\
-1-2^{n+k},&{\rm occurs}\,\frac{2^{n-k}-1}{2^{2k}-1}\,\,\,{\rm
times.}
\end{array}\right.$$
\end{theorem}

\begin{proof} By Equality (\ref{rho1}), we have
$$C_d(\tau)+1=\widehat{\rho_a}(0).$$

From Proposition \ref{pro5}, the rank of $\rho_a$ is $m$ or $m-2k$.
By this together with Lemma \ref{lem1}, one can conclude that
$\widehat{\rho_a}(0)$ takes values from $\{0,
\pm2^{n},\pm2^{n+k}\}$, and $C_d(\tau)$ possibly takes values $-1$,
$-1\pm2^{n}$, $-1\pm2^{n+k}$.

Suppose that $C_d(\tau)$ takes values $-1$, $-1+2^{n}$, $-1-2^{n}$,
$-1+2^{n+k}$ and $-1-2^{n+k}$ exactly $M_1$, $M_2$, $M_3$, $M_4$ and
$M_5$ times, respectively. Obviously,
\begin{equation}\label{equation12}M_1+M_2+M_3+M_4+M_5=2^n-1.\end{equation}
Since $\widehat{\rho_a}(0)$ takes values $\pm2^{n}$ if and only if
the rank of $\rho_a(x)$ is $m$, by Proposition \ref{pro5}, we have
\begin{equation}\label{equation15}M_2+M_3=R_m=\frac{2^{n+2k}-2^{n+k}-2^n+1}{2^{2k}-1}.\end{equation}
By Lemma \ref{lem6} and Proposition \ref{pro6}, we have
\begin{equation}\label{equation11}\left\{\begin{array}{l}
 -M_1+(-1+2^{n})M_2+(-1-2^{n})M_3+(-1+2^{n+k})M_4+(-1-2^{n+k})M_5=1, \\
 2^{2n}M_2+2^{2n}M_3 +2^{2(n+k)}M_4+ 2^{2(n+k)}M_5=2^m(2^n-1), \\
 2^{3n}M_2-2^{3n}M_3+2^{3(n+k)}M_4-2^{3(n+k)}M_5=-2^{2m}+(2^k-2+3)2^{n+m}.
 \end{array}\right.
\end{equation}

The above Equations (\ref{equation12}), (\ref{equation15}) and
(\ref{equation11})
 give
$$\begin{array}{l}
  M_1=2^{n-k}-1,\,\,\,
  M_2=\frac{(2^{n}+1)2^{k-1}}{2^{k}+1},\,\,\,
  M_3=\frac{(2^n-1)(2^{k-1}-1)}{2^{k}-1},\,\,\,
  M_4=0,\,\,\,
  M_5=\frac{2^{n-k}-1}{2^{2k}-1}.
\end{array}
$$

\end{proof}

\begin{remark} It is easy to check that $M_1, M_2$ and $M_5$ are
always nonzero and $M_3=0$ if and only if $k=1$. Thus, when ${\rm
gcd}(l,n)=1$, the cross correlation $C_d(\tau)$ takes exactly three
values, and when ${\rm gcd}(l,n)>1$, the cross correlation takes
exactly four values.
\end{remark}


\begin{remark} By a complete computer experiments for up to $m=26$,
Helleseth et al listed all decimation $d$ such that $C_d(\tau)$
having at most a
six-valued cross-correlation function in \cite{NH4}. 
Their theoretical proofs in \cite{NH4,NH10,HKN6} completely
explained all decimation $d$ listed in \cite{NH4} and giving a
three-valued cross correlation, and they conjectured that the $d$
satisfying $d(2^l+1)=2^i({\rm mod\,} 2^n-1)$ and ${\gcd}(l,n)=1$ are
all decimations that give three-valued cross correlation.

For the case of four-valued cross correlation, only partial classes
of decimations $d$ are shown to have this property \cite{NH11,HKJ}.
The above theorem 1 can explain all decimation $d$ listed in
\cite{NH4} and giving exactly a four-valued cross correlation,
except an exceptional $d=7$ for a small $m=8$ case.

The extensive numerical experiments of Helleseth et al (up to $m=26$
in \cite{NH4}) may suggest one to conjecture, as Helleseth et al do
in \cite{HKN6}, that the converse of Theorem 1 holds true, that is,
for $m\geq 12$, if $C_d(\tau)$ has at most a four-valued cross
correlation, then there must exist some $l$ and $i$ such that
$d(2^l+1)=2^i ({\rm mod\,} 2^n-1)$. If this conjecture can be shown
true, then the conjecture in \cite{HKN6} is also true. This will be
an interesting open problem.
\end{remark}

\begin{remark} When $m$ is a power of 2, there are no $d$ and $l$
satisfying $d(2^l+1)=2^i ({\rm mod\,} 2^n-1)$ by Lemma 1. This
together with the conjecture mentioned above (if true) explains why
there is no $d$ giving at most a four-valued cross correlation for
$m=16$.
\end{remark}

\section{Conclusion}
We have studied the cross correlation between an $m$-sequence of
period $2^m-1$ and the $d$-decimation of an $m$-sequence of period
$2^n-1$ for even $m=2n$. We prove that if $d$ satisfies
$d(2^l+1)=2^i\,\,({\rm mod}\,\, 2^n-1)$ for some $l$ and $i$, then
the cross correlation takes exactly either three or four values,
depending on ${\rm gcd}(l,n)$ is equal to or larger than 1. The
result enriches the work of Helleseth et al. An interesting open
problem is left.

\end{document}